\documentclass[aps,prl,twocolumn,groupedaddress,amsmath,amssymb,nofootinbib,raggedbottom]{revtex4-2}
\usepackage{mathrsfs}
\usepackage{graphicx}
\usepackage{amsmath}
\usepackage{amsfonts}
\usepackage{amssymb}
\usepackage{color}
\usepackage{hyperref}
\usepackage{epsfig}
\usepackage{CJK}
\usepackage{eepic}
\usepackage{bbm}
\usepackage{dcolumn}
\usepackage{bm}
\usepackage{ulem}
\usepackage{multirow}

\newcommand{\omits}[1]{}
\newcommand{\absection}[1]{{\textcolor{black}{\it #1.---}}}

\def\bc{\begin{center}}

\def\ec{\end{center}}
\def\be{\begin{eqnarray}}
\def\ee{\end{eqnarray}}

\definecolor{dyellow}{rgb}{1.,0.8,.0}
\definecolor{myblue}{rgb}{.1,.1,.7}
\definecolor{dcyan}{rgb}{.0,.6,.6}
\definecolor{cyan}{rgb}{0.4,1.0,1.0}
\definecolor{dmagenta}{rgb}{0.6,0.0,0.6}
\definecolor{brown}{rgb}{0.6,0.2,0.}
\definecolor{darkblue}{rgb}{.0,.0,0.5}
\definecolor{darkred}{rgb}{0.75,0.0,0.0}
\definecolor{orange}{rgb}{1.,.6,.0}
\definecolor{dorange}{rgb}{0.8,.4,.0}
\definecolor{green}{rgb}{0.0,1.0,0.0}
\definecolor{darkgreen}{rgb}{0.0,0.6,0.0}
\definecolor{purple}{rgb}{.4,.0,.4}
\definecolor{lightgrey}{rgb}{0.7, 0.7, 0.7}
\definecolor{grey}{rgb}{0.4, 0.4, 0.4}
\definecolor{tiNaLiGreen}{rgb}{0.1, 0.8, 0.2}

\def\be{\begin{equation}}
\def\ee{\end{equation}}
\def\bea{\begin{eqnarray}}
\def\eea{\end{eqnarray}}
\def\e{\epsilon}

\def\o{\omega}

\def\>{\rangle} 
\def\<{\langle} 

\def\ncs{\left(\frac{\mathcal{N}^{(0)}}{c_{\rm s}}\right)}
\def\csn{\left(\frac{c_{\rm s}}{\mathcal{N}^{(0)}}\right)}

\hypersetup{
    colorlinks=true,
    linkcolor=blue,
    citecolor=red,
    urlcolor=black
}

\begin{document}


\title{Analogue Schwinger Pair Production in Acoustic Black Holes}

\author{Ruo-Han Wang} \email{wangrh67@mail2.sysu.edu.cn}
\author{Yu-Ye Cheng} \email{chengyy53@mail2.sysu.edu.cn}
\author{Jia-Rui Sun} \email{sunjiarui@sysu.edu.cn}

\affiliation{School of Physics and Astronomy, Sun Yat-Sen University, Guangzhou 510275, China}



\begin{abstract}
We show that an analogue Schwinger pair production can be realized in Reissner-Nordstr{\"o}m-like acoustic black holes constructed from a Bose-Einstein condensate subject to a synthetic vector potential. Upon Kaluza-Klein reduction of the five-dimensional acoustic geometry, the higher-dimensional phonon modes appear as effectively charged phonons in the resulting four-dimensional spacetime, in which the electric field experienced by the charged phonons arises from the component of the synthetic vector potential along the compactified direction. Then by appropriately choosing the background flow velocity and the synthetic vector potential, the reduced acoustic metric can take the form of a Reissner-Nordstr{\"o}m-like black hole. From this metric we obtain the mean number of spontaneously produced phonon pairs in both the extremal and near extremal cases, and show that in the extremal limit this mean number takes the form of the Schwinger formula. Moreover, charged-phonon pair production not only admits the same effective thermal interpretation as Schwinger pair production of real charged particles, but can also be interpreted as the analogue Hawking radiation and superradiance from the five-dimensional perspective. Finally, we make estimates of pair production using experimental parameters in Bose-Einstein condensate, providing theoretical guidance for future laboratory detection.

\end{abstract}


\maketitle
\newpage
\absection{Introduction}
The notion of acoustic black holes dates back to Unruh’s work~\cite{Unruh:1980cg}, which demonstrated that an acoustic horizon can form in an inviscid, irrotational fluid flow when the velocity exceeds the local sound speed. In this background, the dynamics of acoustic perturbations are governed by the Klein-Gordon (KG) equation for a massless scalar field in curved geometry~\cite{Visser:1997ux}. One of the main motivations for studying analogue gravity is to simulate Hawking radiation in the laboratory, since direct detection of it from black holes remains very difficult. Bose-Einstein condensates (BEC) provide an ideal platform for realizing acoustic black holes~\cite{Garay:1999sk,Barcelo:2000tg,Lahav:2009wx} and acoustic de sitter spacetime~\cite{Fedichev:2003id,Fischer:2004bf}, where analogue Hawking radiation~\cite{Steinhauer:2015saa} with an approximately thermal spectrum~\cite{MunozdeNova:2018fxv} has been observed. Beyond analogy, it was shown that acoustic black holes can correspond to real black holes at the dynamical level, based on the fluid/gravity duality, in which the analogue Hawking temperature of the former can be connected to the Hawking temperature of the latter~\cite{Ge:2015uaa}. Recent work has also explored the information paradox in analogue gravity systems~\cite{Cheng:2025bnj}. For a review on analogue gravity, see~\cite{Barcelo:2005fc} and references therein.

Apart from the Hawking radiation, charged black holes can also emit charged particle pairs through the Schwinger effect, which refers to the creation of charged pairs from vacuum instability induced by a strong electromagnetic field~\cite{Schwinger:1951nm}. The spontaneous pair production of charged particles in charged black holes has been extensively investigated recently, see, for example~\cite{Chen:2012zn, Chen:2014yfa,Kim:2016dmm, Chen:2016caa, Chen:2020mqs, Cai:2020trh,Lin:2024jug, Castellano:2025yur, Chen:2025xrv, Klein:2021les}. Therefore an interesting question is can an analogue or acoustic Schwinger effect also occur in analogue gravity? The essential point is the existence of effectively charged phonons that propagate in a charged acoustic geometry. Several approaches have been proposed to obtain effectively charged phonons, for example, via electron–phonon coupling~\cite{Rice:1976organic, Kuzmenko2009, Cappelluti2012}, a two-component BEC with Raman coupling~\cite{Leizerovitch:2016iwv}, or the Kaluza–Klein (KK) reduction~\cite{Butera2019}. However, the charged acoustic geometry in which charged phonons propagate, and in particular the quantum effects within it, have not been studied so far.

In this article, we apply the KK reduction~\cite{Overduin:1997sri} in a five-dimensional acoustic black hole in a BEC to obtain charged phonons obeying the charged KG equation in the reduced four-dimensional charged acoustic black hole. We show that, for acoustic black holes, the KK reduction extends beyond the generation of effective electromagnetic fields in lower dimensions: it can produce effectively charged phonons and provide charged background geometry in which an analogue Schwinger pair production can occur. Furthermore, compact geometries required by the KK reduction have already been realized in the BEC experiments, such as tunable annular traps supporting persistent flows~\cite{Herve2021} and synthetic cylindrical surfaces subject to synthetic magnetic fluxes~\cite{Li:2022pkt}, which also suggests that our model may be experimentally detectable.
 
In our construction, effectively charged phonons are obtained from the KK reduction of a five-dimensional acoustic black hole with a synthetic vector potential. Although atoms in a BEC are electrically neutral, synthetic vector potentials can make them behave as if they were coupled to an electromagnetic gauge field~\cite{Dalibard:2010ph}.\footnote{Such potentials can be generated using static fields together with microwave or optical Raman beams~\cite{Lin:2008uuz, Lin:2009vfu, Spielman2009}. In acoustic black holes, the synthetic vector potential breaks the irrotational condition and generates an ergoregion, providing a simple setting in which to investigate the superradiant phenomena~\cite{Giacomelli:2020evu}.} The synthetic vector potential plays an additional important role: its component along the compactified direction becomes the gauge potential experienced by the charged phonons in the reduced acoustic geometry. A suitable background flow velocity and external potential, determined from the background equations, then support an RN-like acoustic black hole. Consequently, spontaneous pair production of charged phonons—in particular, the Schwinger effect—can be successfully realized. Explicitly, the mean number of produced charged-phonon pairs is obtained in both the extremal and near-extremal cases, and the thermal interpretation of the acoustic Schwinger effect is given both in four dimensions, following Refs.~\cite{Kim:2015kna, Kim:2015qma}, and in five dimensions. Moreover, using parameters drawn from relevant BEC experiments, we estimate the mean number of produced charged phonons in the extremal RN-like acoustic black hole and find no strong suppression, which offers a possible route toward investigating the Schwinger effect in experimentally accessible analogue gravity systems.

\absection{Acoustic black hole with a synthetic vector potential}
\label{Construction}
In the presence of a synthetic vector potential $\vec{\mathcal{A}}$, the dynamics of the BEC are described by the Gross-Pitaevskii equation with minimal coupling
\be \label{GP}
i \hbar \partial_0 \Psi = \left( \frac{\big(-i \hbar \nabla - \vec{\mathcal{A}}\big)^2}{2M}+V+\lambda |\Psi|^2 \right)\Psi,
\ee
where $V$ is the external potential and $\lambda$ is the coupling constant. Substituting the ansatz of wavefunction $\Psi=\sqrt{\mathcal{N}}\exp (i \theta)$ into Eq.~(\ref{GP}), the real and imaginary parts are, respectively,
\begin{align}
&\hbar\partial_{0}\theta+\frac12 M v_i^2+V+\lambda \mathcal{N}-\frac{\hbar^2}{2M}\frac{\nabla^2 \sqrt{\mathcal{N}}}{\sqrt{\mathcal{N}}}=0, \label{Euler equation}\\
&\partial_{0} \mathcal{N}+\partial_{i}(\mathcal{N}v_i)=0.
\label{continuity}
\end{align}
In the low-momentum limit the last quantum pressure term in Eq.~(\ref{Euler equation}) is neglected. The physical velocity is defined as 
\be \label{physical velocity}
v_i=\frac{\hbar \partial_{i} \theta-\mathcal{A}_i}{M}=v^{\rm c}_i-\frac{\mathcal{A}_i}{M},
\ee
where the canonical velocity $v^{\rm c}_i=\hbar \partial_{i} \theta / M$ satisfies the irrotationality condition. The synthetic vector potential $\vec{\mathcal{A}}$ makes the physical velocity no longer irrotational and provides a route to generating the electric field for phonons in the reduced theory. Although a suitable background flow with nonzero vorticity can also generate such a field, the synthetic vector potential offers a convenient way of tuning its strength. Expanding the hydrodynamic equations perturbatively on a stationary background $\mathcal{N}\to \mathcal{N}^{(0)}+\mathcal{N}^{(1)}, \theta \to \theta^{(0)}+\theta^{(1)}$, we obtain  
\begin{align}
&\lambda \mathcal{N}^{(1)}+\hbar  v_i\partial_i \theta^{(1)}+\hbar \partial_0 \theta^{(1)}=0, \label{Euler equation1} \\
& M\partial_i (\mathcal{N}^{(1)} v_i)+\hbar\partial_i (\mathcal{N}^{(0)} \partial_i \theta^{(1)})+M\partial_0 \mathcal{N}^{(1)}=0,  \label{continuity1}
\end{align}
Eliminating the $\mathcal{N}^{(1)}$ from Eqs.~(\ref{Euler equation1})(\ref{continuity1}) gives the massless KG equation for $\theta^{(1)}$
\be \label{massless KG eq}
\frac{1}{\sqrt{-G}}\partial_A\left(\sqrt{-G}G^{AB}\partial_B \theta^{(1)}\right)=0.
\ee
with
\be \label{(4+1)d metric}
G_{AB}=\ncs^\frac 23
\begin{pmatrix}
-(c_{\rm s}^2-v^2) & -v_j \\
-v_i               &  \delta_{ij}
\end{pmatrix}
\ee
where $c_{\rm s}^2=\lambda \mathcal{N}^{(0)} / M$ is the local sound speed. Here, uppercase Latin indices ($A, B= 0, 1, 2, 3 ,4$) denote the five-dimensional spacetime and lowercase Latin indices ($i= 1, 2, 3, 4$) correspond to the spatial dimensions.

The synthetic vector field allows the neutral condensate to behave as a charged fluid acted upon by electromagnetic fields. This interpretation becomes explicit upon rearranging Eq.~(\ref{Euler equation}) into the standard Euler form for a charged fluid with unit charge
\be
M\left[ \partial_0\vec{v}+ (\vec{v}\cdot \nabla)\vec{v}\right]=\mathfrak{q} (\vec{E}_{\mathrm{syn}}+\vec{v} \times \vec{B}_{\mathrm{syn}}) -  \nabla(V+\lambda \mathcal{N}- \Phi ) ,
\ee
with $\mathfrak{q}=1$ and $\Phi$ is the electric potential acting on the fluid. The synthetic vector field then acts as gauge field couples to the neutral atoms as if they carried an effective charge. The effective charge of the phonons, however, emerges through the KK reduction. In contrast, a charged atoms background only modify the acoustic geometry experienced by phonons~\cite{Anacleto:2010cr, Ge:2010wx, Cropp:2015tua}. 

\absection{Kaluza-Klein reduction}
\label{kk}
To obtain the charged phonons in four-dimensional acoustic black holes from the KK reduction of five-dimensional acoustic black holes, we introduce coordinates $x^A=(t, x^a, y)$ for the background manifold $\mathbb{R}^{1, 3}\times S^1$, where $a=1, 2, 3$ and $y \sim y +2\pi R_{\rm c}$ with $R_{\rm c}$ denoting the compactification radius. The background condensate density $\mathcal{N}^{(0)}$, sound speed $c_{\rm s}$, velocity $v_i$, and synthetic gauge potential $\mathcal{A}_i$ are assumed to be independent of the compact coordinate $y$. Using Eq.~(\ref{(4+1)d metric}), we can rewrite the five-dimensional line element as
\begin{align}
ds^2=&\ncs^\frac 23\left[-(c_{\rm s}^2-v^2)dt^2-2v_a dx^a dt+\delta_{ab}dx^a dx^b\right]\cr
&+\ncs^\frac 23\left(dy-v_y dt\right)^2.
\end{align}
Starting from the standard ansatz for the five-dimensional KK line element~\cite{Overduin:1997sri}
\be
ds^2=\chi^{-\frac 13}g_{\mu\nu}dx^\mu dx^\nu+\chi^{\frac 23}(dy+A_\mu dx^\mu)^2.
\ee
One can directly read off the four-dimensional acoustic metric $g_{\mu\nu}$ (Greek letters represent the indices of the four-dimensional acoustic spacetime coordinates), the $U(1)$ gauge field $A_\mu=(-v_y, 0, 0, 0)$ and the dilaton $\chi=\mathcal{N}^{(0)}/c_{\rm s}$. The periodicity $y\sim y+2\pi R_{\rm c}$ allows a Fourier expansion of the phonon field,
\be \label{Fourier expansion}
\theta^{(1)}(x^\mu, y)=\frac{1}{\sqrt{2\pi R_{\rm c}}}\sum_{n=-\infty}^{\infty}\phi_n(x^\mu)e^{i q_n y},
\ee
with $q_n=n/R_{\rm c}$ being the charge carried by the $n$-th phonon mode. Therefore, the massless KG equation of five-dimensional phonons, i.e. Eq.~(\ref{massless KG eq}) can be reduced to the four-dimensional KG equation for massive complex scalar fields
\be \label{charged KG eq}
D_\mu D^\mu \phi_n - \csn q_n^2 \phi_n=0
\ee
with $D_\mu= \nabla_\mu-i q_n A_\mu$ and mass square $\mu_n^2=\csn q_n^2$, as well as charge $ q_n$ for the $n$-th phonon mode $\phi_n$. Alternatively, by substituting Eq.~(\ref{Fourier expansion}) into the action of the five-dimensional phonon $\theta^{(1)}$
\be
S=-\frac 12 \int d^4x dy \sqrt{-G}G^{AB}\partial_A\theta^{(1)} \partial_B \theta^{(1)},
\ee
one obtains that the $n$-th phonon mode $\phi_n$ is governed by the action 
\be \label{charged phonon action}
S_n=-\frac 12 \int d^4x \sqrt{-g} \left[ D^*_\mu \phi^*_n D^\mu \phi_n+\csn q_n^2 \phi^*_n \phi_n \right]
\ee
with $S=\sum_n S_n$. 

\absection{Reissner-Nordstr{\"o}m-like acoustic black hole}
Choosing the radial flow speed as $v(r)=c_{\rm s}\sqrt{2M_{\rm s}/r-Q_{\rm s}^2/r^2}$~\cite{Liberati:2000pt} can yield a four-dimensional RN-like acoustic black hole
\be \label{acoustic RN metric}
ds^2=\frac{\mathcal{N}^{(0)}}{c_{\rm s}}\left[- c_{\rm s}^2 f_{\mathrm{RN}}(r)dt^2+\frac{dr^2}{f_{\mathrm{RN}}(r)}+r^2d\Omega_2^2\right]. 
\ee
Here
\be
f_{\mathrm{RN}}(r)=1-\frac{2M_{\rm s}}{r}+\frac{Q_{\rm s}^2}{r^2}=\left(1-\frac{r_+}{r}\right)\left(1-\frac{r_-}{r}\right)
\ee
with $r_++r_-=2M_{\rm s}$ and $r_+r_-=Q_{\rm s}^2$, where $M_{\rm s}$ and $Q_{\rm s}$ act as the mass and charge of the RN geometry, respectively. Then denoting the acoustic-horizon radius $r_+=r_{\rm h}=r_-/\eta$, the sound speed at the acoustic horizon to be $c_{\rm h}$ and $\mathcal{N}^{(0)}(r_{\rm h})=\mathcal{N}^{(0)}_{\rm h}$, see~\cite{suppl} for details.
\omits{
\begin{alignat}{2}
r_+&=r_{\rm h},                                                                  &        \qquad  \quad   \qquad               r_-&=\eta r_{\rm h}, \cr
M_{\rm s}&=\frac{1+\eta}{2}r_{\rm h},                                     &        \qquad  \quad  \qquad              Q_{\rm s}&=\sqrt{\eta}r_{\rm h},  \cr
\mathcal{N}^{(0)}_{\rm h}&=\frac{c_{\rm h}^2 M}{\lambda},          &        \qquad  \quad  \qquad                   J&=\frac{M r_{\rm h}^2 c_{\rm h}^3}{\lambda}  .
\end{alignat} 
$\mathcal{N}^{(0)}_{\rm h}$ is the condensate density at the horizon. 
}

\absection{Spontaneous pair production}
\label{pp}
Spontaneous pair production in a RN black hole takes place primarily in the near horizon region~\cite{Chen:2012zn}, which acts as the causal boundary for Hawking radiation and the region of electric field domination for the Schwinger effect. We mainly follow~\cite{Chen:2012zn} and take the near horizon and the near extremal limits
\begin{equation} \label{rescal}
r \to Q_{\rm s} + \epsilon \rho, \qquad M_{\rm s} \to Q_{\rm s} + \frac{\epsilon^2 B^2}{2 Q_{\rm s}}, \qquad t \to \frac{\tau}{c_{\rm h} \epsilon},
\end{equation}
The parameter $B$ serves as the new horizon radius for the new black hole and it can be expressed as $B=\sqrt{\sqrt{\eta}(1+\eta-2\sqrt{\eta})}/r_{\rm h} \e$ and $\e$ is an infinitesimal parameter. Since the condensate density and the speed of sound vary slowly over such a narrow region, they can be treated as constants. We choose the compact component of the synthetic vector potential to be $\mathcal{A}_y=k M Q_{\rm s} c_{\rm h} / r $, which gives rise to the electric field experienced by the phonons after the KK reduction.\footnote{The charged phonons couple to the gauge potential $A_\mu=(-v_y, 0, 0, 0)$, where $v_y=v_{y}^{\rm c}-\mathcal{A}_y / M$, as given by Eq.~(\ref{physical velocity}). The irrotationality condition requires $\partial_i v_{j}^{\rm c} - \partial_j v_{i}^{\rm c} =0$. Setting $ j=y$ and using $\partial_y v_{i}^{\rm c}=0$ (The background is independent of the compact coordinate y) we obtain $\partial_i v_{y}^{\rm c}=0$. Hence the electric field experienced by the phonons is determined by the synthetic vector potential.} Here $k$ is a constant controlling the strength of the synthetic vector potential. With this choice of synthetic vector potential, the phonons experience an effective spherically symmetric electric field as if it is generated by a point charge $kQ_s$ at the center of the black hole. We can therefore control the effective charge of the acoustic black hole by tuning the synthetic vector potential. The near horizon near extremal RN-like acoustic metric with a $U(1)$ gauge field is given by
\begin{align}\label{near extremal acoustic RN}
ds^2&=\frac{\mathcal{N}^{(0)}_{\rm h}}{c_{\rm h}}\left[-\frac{\rho^2-B^2}{Q_{\rm s}^2}d\tau^2+\frac{Q_{\rm s}^2d\rho^2}{\rho^2-B^2}+Q_{\rm s}^2d\Omega_2^2\right], \cr
A&=-k\frac{\rho}{Q_{\rm s}}d \tau,     \qquad      F=\frac{k}{Q_{\rm s}}d\tau \wedge d \rho.
\end{align}
The metric~(\ref{near extremal acoustic RN}) differs from the corresponding RN black hole only by a constant conformal factor. Decomposing the scalar field in spherical coordinates as $\phi_n(\tau, \rho, \theta, \varphi)=e^{-i \o \tau + i m \varphi}R(\rho) S(\theta)$, one has
\begin{align} \label{near extremal eq rho}
&\partial_\rho\left[\left(\rho^2-B^2\right)\partial_\rho R\right]+\left[\frac{Q_{\rm s}^2(Q_{\rm s} \o- k q_n \rho)^2}{\rho^2-B^2} \right. \cr
&\left. -q_n^2Q_{\rm s}^2-l(l+1)\right]R=0. 
\end{align}
The condition for creating charged phonon pairs corresponds to violation of the Breitenlohner-Freedman (BF) bound in AdS$_2$ spacetime
\be \label{BF}
Q_{\rm s}^2\left(q_n^2-k^2 q_n^2\right)+\left(l+\frac 12 \right)^2<0,
\ee
which requires $k>1$. For convenience, we define $\tilde{q}_n\equiv k q_n$ and $\bar{\mu}_n\equiv q_n$. $k>1$ indicates that, for radiation to occur, the synthetic vector field must be tuned such that 
$\tilde{q}_n>\bar{\mu}_n$.
\omits{we rewrite eq. (\ref{near extremal eq rho}) as
\begin{align}
&\partial_\rho\left[\left(\rho^2-B^2\right)\partial_\rho R\right]+\left[\frac{Q_{\rm s}^2(\rho \tilde{q}_n-\o Q_{\rm s})^2}{\rho^2-B^2}  \right. \cr
 & \left. -\bar{\mu}_n^2Q_{\rm s}^2-l(l+1)\right]R=0
\end{align}
which describes the motion of the charged phonons with mass $q_m$ and charge $k q_m$ in the near horizon region in the near extremal RN black hole under a gauge potential $\rho d\tau/Q_{\rm s}$.} 

\absection{(A). Extremal case}
The extremal limit of the RN-like acoustic black hole is $B=0$ or $r_{\rm h}=r_-=Q_{\rm s}$, namely, the analogue Hawking temperature vanishes, thus only the analogue Schwinger effect exists. The mean number of produced pairs $|\beta|^2$ can be obtained by comparing the result in~\cite{Chen:2012zn}, which is 
\begin{align}
|\beta|^2 &= \frac{\sinh (2 \pi b)}{\cosh (\pi a + \pi b)} e^{ \pi b - \pi a}, \label{Ebeta}
\end{align}
where
\be
a\equiv \tilde{q}_n Q_{\rm s},  \qquad b \equiv \sqrt{(\tilde{q}_n^2-\bar{\mu}_n^2)Q_{\rm s}^2-\left(l+\frac 12\right)^2}. 
\ee
In the limit $k \gg 1$, the mean number of produced pairs is approximately the Schwinger formula
\begin{align} \label{Schw form}
|\beta|^2 \approx \exp\left(- \frac{\pi \bar{\mu}_n^2 Q_{\rm s}}{\tilde{q}_n}\right)=\exp\left(-\frac{n\pi Q_{\rm s} }{k R_{\rm c}}\right).
\end{align}
Here $n>0$, since the produced phonons carry positive charge $\tilde{q}_n>0$.

Next, we will analyze the thermal interpretation of the acoustic Schwinger effect. Defining the effective mass $\mu_{\mathrm{eff}}$ as
\be
\mu_{\mathrm{eff}}^2 =\left(\frac{c_{\rm h}}{\mathcal{N}^{(0)}_{\rm h}}\right)\left(\bar{\mu}_n^2 +\left(l+\frac 12\right)^2/Q_{\rm s}^2\right).
\ee
The Davies-Unruh temperature associated with the acceleration of a charge near the acoustic horizon is
\be
T_U=\frac{\frac{\tilde{q}_n E}{\mu_{\mathrm{eff}}}}{2\pi}=\frac{\tilde{q}_nc_{\rm h} }{2\pi \mu_{\mathrm{eff}}Q_{\rm s} \mathcal{N}^{(0)}_{\rm h}},
\ee
where $E=c_{\rm h}/(\mathcal{N}^{(0)}_{\rm h} Q_{\rm s})$ denotes the electric field strength measured in a local frame. The effective temperature on the horizon $T_\mathrm{RN}$ and another parameter $\mathcal{T}_\mathrm{RN}$ which is associated with $T_\mathrm{RN}$ are defined as
\begin{align}
T_\mathrm{RN}&=T_U+\sqrt{T_U^2+\frac{\mathcal{R}_{\mathrm{AdS}}}{8\pi^2}}\cr
&=\frac{c_{\rm h}}{2\pi \mu_{\mathrm{eff} }Q_{\rm s} \mathcal{N}^{(0)}_{\rm h}}\left(\tilde{q}_n+\sqrt{\tilde{q}_n^2-\frac{\mu_{\mathrm{eff}}^2\mathcal{N}^{(0)}_{\rm h}}{c_{\rm h}}}\right).\cr
\mathcal{T}_\mathrm{RN}&=T_U-\sqrt{T_U^2+\frac{\mathcal{R}_{\mathrm{AdS}}}{8\pi^2}}\cr
&=\frac{c_{\rm h}}{2\pi \mu_{\mathrm{eff} }Q_{\rm s} \mathcal{N}^{(0)}_{\rm h}}\left(\tilde{q}_n-\sqrt{\tilde{q}_n^2-\frac{\mu_{\mathrm{eff}}^2\mathcal{N}^{(0)}_{\rm h}}{c_{\rm h}}}\right),
\end{align}
where $\mathcal{R}_{\mathrm{AdS}}=-2 c_{\rm h} / (\mathcal{N}^{(0)}_{\rm h} Q_{\rm s}^2$) is the Ricci scalar of the AdS$_2$ in metric~(\ref{near extremal acoustic RN}). Then $|\beta|^2$ can be rewritten as
\be
|\beta|^2 =\frac{e^{2\pi b -2\pi a}-e^{-2\pi b-2\pi a}}{1+e^{-2\pi a-2\pi b}}=\frac{e^{-\frac{\mu_{\mathrm{eff}}}{T_\mathrm{RN}}}-e^{-\frac{\mu_{\mathrm{eff}}}{\mathcal{T}_\mathrm{RN}}}}{1+e^{-\frac{\mu_{\mathrm{eff}}}{\mathcal{T}_\mathrm{RN}}}}.
\ee

Furthermore, considering the strong synthetic field $(\tilde{q}_n\gg \mu_{\mathrm{eff}} \sqrt{\mathcal{N}^{(0)}_{\rm h}/c_{\rm h}} )$ and small angular momentum ($\bar{\mu}_n\gg l$) give
\begin{align}
T^\prime&=T_\mathrm{RN}(\tilde{q}_n\gg \mu_{\mathrm{eff}}\sqrt{\mathcal{N}^{(0)}_{\rm h}/c_{\rm h}} , \bar{\mu}_n\gg l) \cr
&\simeq\sqrt{\frac{c_{\rm h}}{\mathcal{N}^{(0)}_{\rm h}}}\frac{\tilde{q}_n}{\pi \bar{\mu}_nQ_{\rm s}}=
\sqrt{\frac{c_{\rm h}}{\mathcal{N}^{(0)}_{\rm h}}}\frac{k}{\pi r_{\rm h}}. 
\end{align}
For a positive effective background charge, the total number of produced pairs can then be obtained by summing over the positive KK modes as
\be \label{total number}
\sum_{n=1}^{\infty}|\beta|^2 \approx \sum_{n=1}^{\infty}\exp\left(-\frac{n\pi r_{\rm h}}{k R_{\rm c}}\right)=\frac{1}{e^\frac{\pi r_{\rm h}}{k R_{\rm c}}-1}=\frac{1}{e^\frac{1}{ R'_c T^\prime}-1},
\ee
with $R'_c=\sqrt{\frac{\mathcal{N}^{(0)}_{\rm h}}{c_{\rm h}}}R_{\rm c}$. Under these approximations, Eq.~(\ref{total number}) takes a form of the Bose-Einstein distribution, with temperature $T^\prime$ and energy gap $1/R'_c$. 

\absection{(B). Near extremal case}
In the near extremal RN-like acoustic black hole, its analogue Hawking temperature is $T_{\rm H}=B/(2\pi Q_{\rm s}^2)$, the gauge potential $\Phi_{\rm h}=kB/Q_{\rm s}$, and we denote $\tilde{a}= \o Q_{\rm s}^2/B$. Similarly, the mean number of produced pairs can be obtained as
\omits{\begin{align}
|\beta|^2 & = \frac{\sinh (2 \pi b) \sinh (\pi \tilde a - \pi a )}{\cosh (\pi a + \pi b) \cosh (\pi \tilde a - \pi b)}, \label{Nbeta}
\end{align}
whose explicit form takes}
\begin{align} \label{termal interpretation}
 &|\beta|^2  = \frac{e^{2\pi b -2\pi a}-e^{-2\pi b-2\pi a}}{1+e^{-2\pi a-2\pi b}}\  \frac{1-e^{-2\pi \tilde{a}+2\pi a}}{1+e^{-2\pi \tilde{a}+2\pi b}}\cr
&=e^{\frac{\mu_{\mathrm{eff}}}{T_\mathrm{RN}}} \! \!  \times \! \frac{e^{-\frac{\mu_{\mathrm{eff}}}{T_\mathrm{RN}}} \! \! - \! e^{-\frac{\mu_{\mathrm{eff}}}{\mathcal{T}_\mathrm{RN}}}}{1+e^{-\frac{\mu_{\mathrm{eff}}}{\mathcal{T}_\mathrm{RN}}}} \! \! \times \! \frac{e^{-\frac{\mu_{\mathrm{eff}}}{T_\mathrm{RN}}}\left(1 \! -e^{-\frac{\o -q_n\Phi_{\rm h}}{T_{\rm H}}} \! \right)}{1+e^{-\frac{\mu_{\mathrm{eff}}}{T_\mathrm{RN}}}e^{-\frac{\o -q_n\Phi_{\rm h}}{T_{\rm H}}}}, \cr 
\end{align}
which admits a clear thermal interpretation as in the near extremal charged black holes~\cite{Chen:2016caa}: the second term describes the analogous Schwinger effect in the AdS$_2$ with effective temperature $T_\mathrm{RN}$, and the third term is the analogous Schwinger effect in the Rindler space with the Unruh temperature given by the analogous Hawking temperature $T_{\rm H}$. 


\absection{Thermal interpretation from the five-dimension}\label{sup}
Spontaneous pair production in the four-dimensional RN-like acoustic black hole admits a different interpretation from the five-dimensional perspective. Note that in the near horizon near extremal limits, the five-dimensional acoustic metric takes the form 
\begin{align}
d s^2&=\left(\frac{\mathcal{N}^{(0)}_{\rm h}}{c_{\rm h}}\right)^\frac 23 \left[-\frac{\rho^2-B^2}{Q_{\rm s}^2}d\tau^2+\frac{Q_{\rm s}^2d\rho^2}{\rho^2-B^2}+Q_{\rm s}^2d\Omega_2^2 \right]\cr
&+\left(\frac{\mathcal{N}^{(0)}_{\rm h}}{c_{\rm h}}\right)^\frac 23 \left(R_{\rm c}d\varphi-k \frac{\rho}{Q_{\rm s}} d\tau\right)^2,
\end{align}
with angular velocity $\Omega_{\rm h}=kB/(Q_{\rm s}R_{\rm c})$ at the horizon, where $R_{\rm c}d\varphi=dy$. Substituting the metric into Eq.~(\ref{massless KG eq}) reproduces Eq.~(\ref{near extremal eq rho}), which now describes the scattering of neutral massless phonons with angular momentum $n$ along the compact direction. In the five-dimensional acoustic black hole, both the Hawking radiation and the superradiance (with $\omega<n\Omega_{\rm h}$) can exist~\cite{Bredberg:2009pv}. As $\rho\to\infty$, the solution of Eq.~(\ref{near extremal eq rho}) behaves asymptotically as
\be
R(\rho) \sim\rho^{-1/2\pm ib}, \quad b=\sqrt{(k^2-1)Q_{\rm s}^2\frac{n^2}{R^2_{\rm c}}-\left(l+\frac12\right)^2}.
\ee
The existence of propagating modes requires $b$ to be real, i.e., Eq.~(\ref{BF}) is satisfied. And the four-dimensional thermal interpretation in Eq.~(\ref{termal interpretation}) needs to be reformulated in the five-dimensional form as
\begin{align}\label{5d thermal}
|\beta|^2 =e^{\frac{\mu_{\mathrm{eff}}}{T_\mathrm{RN}}} \!\! \times \! \!\!\! \!\underbrace{\frac{e^{-\frac{\mu_{\mathrm{eff}}}{T_\mathrm{RN}}}-e^{-\frac{\mu_{\mathrm{eff}}}{\mathcal{T}_\mathrm{RN}}}}{1+e^{-\frac{\mu_{\mathrm{eff}}}{\mathcal{T}_\mathrm{RN}}}} }_{\substack{\text{Spontaneous  superradiance }\\ \text {in   AdS}_2}} \!\!\!\! \! \! \times 
 \underbrace{\frac{e^{-\frac{\mu_{\mathrm{eff}}}{T_\mathrm{RN}}}\left(1-e^{-\frac{\o -n\Omega_{\rm h}}{T_{\rm H}}}\right)}{1+e^{-\frac{\mu_{\mathrm{eff}}}{T_\mathrm{RN}}}e^{-\frac{\o -n\Omega_{\rm h}}{T_{\rm H}}}}}_{\substack{\text{Hawking $+$ superradiance}\\ \text {in  Rindler  space} }}.
\end{align}
Note that in the extremal limit $T_{\rm H}\to 0$, $|\beta|^2$ in Eq.~(\ref{5d thermal}) is dominated by the superradiance with the absorption cross section ratio $\sigma_{\rm abs}<0$.

\absection{Estimates of pair production using experimentally relevant parameters}
To estimate the expected mean number of produced charged phonon pairs, we consider an extremal acoustic black hole $(\eta=1)$ with a horizon radius of $r_{\rm h}=10\,\mu\mathrm{m}$ and a compactification radius of $R_{\rm c}=20\,\mu\mathrm{m}$. For the former, an acoustic black hole with horizon radius of approximately $10-15\,\mu\mathrm{m}$ has been realized experimentally in a convergent radial flow of a two-dimensional cesium superfluid~\cite{Tamura:2023mby}. The latter lies within the $20-150\,\mu{\rm m}$ range of ring radii achieved in BEC~\cite{Herve2021}. We set the horizon sound speed to $c_{\rm h}=0.5\,\mathrm{mm/s}$, comparable to the exterior value of $0.52\,\mathrm{mm/s}$ measured in an acoustic black hole experiment~\cite{MunozdeNova:2018fxv}. For $k=10$, we use the atomic mass of $^{87}\mathrm{Rb}$ and the required synthetic vector potential at the horizon to be $\mathcal{A}_y(r_{\rm h})\simeq7.2\times10^{-28} $ kg m s$^{-1}$. This value falls within the range $|\mathcal{A}_y(r_{\rm h})|<1.17\times10^{-27} $ kg m s$^{-1}$ of synthetic vector potentials demonstrated in the experiment~\cite{Lin:2009vfu}. For the $s$-wave $(l=0)$ of the lowest nonzero mode ($n=1$), the above parameters give $a=5, b=\sqrt{24.5}$. The exact mean number of produced Schwinger pairs in Eq.~(\ref{Ebeta}) then gives $|\beta_{n=1,l=0}|^2\simeq0.729$ in the extremal case. This dimensionless mean pair number indicates that production in the selected mode is not strongly suppressed, providing a strong motivation for further investigation of experimental detectability.

\absection{Conclusions and outlook}\label{con}
In this paper, we study the analogue Schwinger pair production in an acoustic analogue of a (near) extremal RN black hole. We begin with the construction of a five-dimensional acoustic black hole emerged from the BEC subject to a synthetic vector potential. Through the KK reduction, the effectively charged phonons propagating in the corresponding four-dimensional acoustic geometry are obtained. Then by appropriately choosing the background flow velocity and the synthetic vector potential, we obtain the four-dimensional RN-like acoustic black hole, in which the spontaneously pair production, especially the Schwinger effect, of charged phonons is successfully realized. In addition, the acoustic Schwinger effect is shown to admit an effective thermal interpretation parallel to that of genuine Schwinger pair production by real charged particles. Furthermore, the thermal interpretation can be understood in the five-dimensional acoustic geometry as the Hawking radiation and the superradiance. Moreover, we make estimates of pair production by using experimental parameters in BEC system, which provides motivation for future investigation of experimental detectability.

Several questions are left open for future study. In our construction, charged phonons naturally experience an electric field after the KK reduction. It is of interest to explore whether charged phonons coupled to a magnetic field can likewise be constructed in the resulting acoustic black hole. Moreover, a synthetic vector potential is not the only means of generating an effective electric field for charged phonons. It would therefore be worthwhile to investigate alternative mechanisms for generating such a field. As AdS$_2$ spacetime provides a well established setting for Schwinger pair production, a natural extension is to investigate its acoustic analogue. 
\paragraph*{Acknowledgements---}
This work was supported by the National Natural Science Foundation
of China (No.~12475069) and the Guangdong Basic and Applied Basic Research
Foundation (No.~2025A1515011321).




\onecolumngrid
\newpage
\begin{center}
{\Large\bfseries \textemdash{} Supplemental Material \textemdash{}\par}
\vspace{0.8em}
{\Large\bfseries Analogue Schwinger Pair Production in Acoustic Black Holes\par}
\vspace{1em}
{\large Ruo-Han Wang, Yu-Ye Cheng, and Jia-Rui Sun\par}
\end{center}
\setcounter{equation}{0}
\begin{appendix}

\section{Appendix A}
In this Appendix, we present the construction of acoustic Reissner-Nordstr{\"o}m-like black hole. If the radial flow speed is chosen as $v(r)=c_{\rm s}\sqrt{2M_{\rm s}/r-Q_{\rm s}^2/r^2}$, the acoustic metric takes the form
\be \label{acoustic RN metric}
ds^2=\frac{\mathcal{N}^{(0)}}{c_{\rm s}}\left[- c_{\rm s}^2 f_{\mathrm{RN}}(r)dt^2+\frac{dr^2}{f_{\mathrm{RN}}(r)}+r^2d\Omega_2^2\right]. 
\ee
Here
\be
f_{\mathrm{RN}}(r)=1-\frac{2M_{\rm s}}{r}+\frac{Q_{\rm s}^2}{r^2}=\left(1-\frac{r_+}{r}\right)\left(1-\frac{r_-}{r}\right)
\ee
with $r_++r_-=2M_{\rm s}$ and $r_+r_-=Q_{\rm s}^2$. $M_{\rm s}$ and $Q_{\rm s}$ are parameters introduced to reproduce the RN geometry. To realize this flow, we allow the sound speed and density to vary with position and require the following equations to hold simultaneously
\begin{align}
&\frac{1}{r^2}\frac{d}{dr}\left(r^2\mathcal{N}^{(0)}(r)v(r)\right)=0, \cr 
&c_{\rm s}^2(r)=\frac{\lambda \mathcal{N}^{(0)}(r)}{M}, \cr
& v(r)=c_{\rm s}(r)\sqrt{\frac{2M_{\rm s}}{r}-\frac{Q_{\rm s}^2}{r^2}}.
\end{align}
They can be solved as follows:
\begin{align}
c_{\rm s}(r)&=\left(\frac{J\lambda}{M r^2}\right)^\frac 13\left(\frac{2M_{\rm s}}{r}-\frac{Q_{\rm s}^2}{r^2}\right)^{-\frac{1}{6}},\cr
\mathcal{N}^{(0)}(r)&=\left(\frac{J^2 M}{r^4 \lambda}\right)^\frac 13 \left(\frac{2M_{\rm s}}{r}-\frac{Q_{\rm s}^2}{r^2}\right)^{-\frac{1}{3}},\cr
v(r)&=\left(\frac{J \lambda}{M r^2}\right)^\frac 13 \left(\frac{2M_{\rm s}}{r}-\frac{Q_{\rm s}^2}{r^2}\right)^{\frac{1}{3}}.
\end{align}
where $J=r^2\mathcal{N}^{(0)}(r)v(r)=\mathrm{const.}$. The potential required to produce this flow is
\begin{equation}
V=-\left(\frac{J^2 M \lambda^2}{r^4}\right)^\frac 13 \left(\frac{2 M_{\rm s}}{r}-\frac{Q_{\rm s}^2}{r^2}\right)^{-\frac 13} \left(\frac{M_{\rm s}}{r}-\frac{Q_{\rm s}^2}{2r^2}+1 \right).
\end{equation}
All the quantities introduced above, such as $J$, $M_{\rm s}$, and $Q_{\rm s}$, can be expressed in terms of five independent quantities: the mass of the constituent atoms $M$, the coupling constant $\lambda$, the acoustic horizon radius $r_{\rm h}$, the sound speed at the acoustic horizon $c_{\rm h}$ and $\eta=r_-/r_+$, which characterizes the geometric profile of the RN-like acoustic black hole.
\begin{alignat}{2}
r_+&=r_{\rm h},                                                                  &        \qquad  \quad   \qquad               r_-&=\eta r_{\rm h}, \cr
M_{\rm s}&=\frac{1+\eta}{2}r_{\rm h},                                     &        \qquad  \quad  \qquad              Q_{\rm s}&=\sqrt{\eta}r_{\rm h},  \cr
\mathcal{N}_{\rm h}^{(0)}&=\frac{c_{\rm h}^2 M}{\lambda},          &        \qquad  \quad  \qquad                   J&=\frac{M r_{\rm h}^2 c_{\rm h}^3}{\lambda}  .
\end{alignat} 
$\mathcal{N}_{\rm h}^{(0)}$ is the condensate density at the horizon. 

\section{Appendix B}
We can utilize the result in~\cite{Chen:2012zn} to get the vacuum persistence amplitude $|\alpha|^2$, the mean number of pairs $|\beta|^2$ and the absorption cross section $\sigma_\mathrm{abs}$ of the charged phonons. In the extremal acoustic RN black hole, the Hawking temperature vanishes ($B$=0), so that only the analog Schwinger effect exists. We define $z=2 i \omega Q_s^2/\rho$ and the KG equation can be expressed as the Whittaker equation
\begin{equation}
\left(\partial_z^2-\frac 14 + \frac{i a}{z}+\frac{\frac14 -(ib)^2}{z^2}\right)R(z)=0,
\end{equation}
where
\be
a\equiv \tilde{q}_n Q_{\rm s},  \qquad b \equiv \sqrt{(\tilde{q}_n^2-\bar{\mu}_n^2)Q_{\rm s}^2-\left(l+\frac 12\right)^2}. 
\ee
The general solution for the radial equation can be expressed in terms of the Whittaker function $M_{\kappa,\mu}(z)$ as
\be
R(z) = c_1 M_{ia,-ib}(z) + c_2 M_{ia,ib}(z).
\ee

The flux of a charged phonon mode
$\phi_n$ is
\begin{equation}
D=i\,\sqrt{-g}\,g^{\rho\rho}
\left[\phi_n(D_\rho\phi_n)^*-\phi_n^*D_\rho\phi_n\right],
\end{equation}

We impose the inner boundary condition at the extremal
acoustic horizon ($\rho=0$), so that there is no outgoing flux at the inner boundary $D_H^{(\mathrm{out})}=0$. At the outer boundary of the near-horizon
region ($\rho\to\infty$), the ingoing flux
$D_B^{(\mathrm{in})}$ is incident on the scattering region, and
the outgoing flux $D_B^{(\mathrm{out})}$ is reflected back toward the boundary. The transmitted flux $D_H^{(\mathrm{in})}$ crosses the horizons. Evaluating the corresponding fluxes
\begin{align}
&D_B^\mathrm{(in)} = - 2 \omega Q^2 \left( 2 b \, e^{- \pi b} \right) \left| c_1 \right|^2, \qquad  \cr &D_H^\mathrm{(in)} = - 2 \omega Q^2 \, e^{\pi a} \left| c_H^\mathrm{(in)} \right|^2, \cr
&D_B^\mathrm{(out)} = 2 \omega Q^2 \left( 2 b \, e^{\pi b} \right) \left| c_2 \right|^2, \qquad  \cr
&D_H^\mathrm{(out)} = 0 ,
\end{align}
where  
\begin{align}
c_H^\mathrm{(in)} = c_1 e^{-\pi(a + b)} \frac{\sinh(2 \pi b)}{\cosh(\pi a - \pi b)} \frac{\Gamma\left( 1 + 2ib \right)}{\Gamma\left( \frac12 - ia + ib \right)},
\end{align}
and
\begin{equation}
c_2 = - c_1 \frac{\Gamma\left( 1 + 2ib \right) \Gamma\left( \frac12 + ia - ib \right)}{\Gamma\left( 1 - 2ib \right) \Gamma\left( \frac12 + ia + ib \right)} e^{-2\pi b}.
\end{equation}
The Bogoliubov coefficients $\alpha$, $\beta$ and the absorption cross section $\sigma_\mathrm{abs}$ can be obtained
\begin{align}
|\alpha|^2 &=\frac{|D_B^{(\mathrm{in})}|}{|D_B^{(\mathrm{out})}|}= \frac{\cosh (\pi a - \pi b)}{\cosh (\pi a + \pi b)} e^{2 \pi b}, \label{Ealpha}
\\
|\beta|^2 &=\frac{|D_H^{(\mathrm{in})}|}{|D_B^{(\mathrm{out})}|}= \frac{\sinh (2 \pi b)}{\cosh (\pi a + \pi b)} e^{ \pi b - \pi a}, \label{Ebeta}
\\
\sigma_\mathrm{abs} &= \frac{|\beta|^2}{|\alpha|^2} =\frac{\sinh(2 \pi b)}{\cosh(\pi a - \pi b)}e^{- \pi a - \pi b} . \label{Esigma}
\end{align}
In the near extremal case, we define
\be
\tilde{a}\equiv \frac{\o Q_{\rm s}^2}{B}. 
\ee
The solution for the KG equation takes the form of the hypergeometric function
\begin{align}
R(\rho) =& c_1 (\rho - B)^{- \frac{i}2 (a - \tilde a)} (\rho + B)^{\frac{i}2 (a + \tilde a)} \, F\left( \frac12 + i \tilde a + i b,  \frac12 + i \tilde a - i b; 1 - i a + i \tilde a; \frac12 - \frac{\rho}{2B} \right)
\nonumber\\
+& c_2 (\rho - B)^{\frac{i}2 (a - \tilde a)} (\rho + B)^{\frac{i}2 (a + \tilde a)} \, F\left( \frac12 + i a + i b,  \frac12 + i a - i b; 1 + i a - i \tilde a; \frac12 - \frac{\rho}{2B} \right).
\end{align}
Similarly, we can obtain
\begin{align}
|\alpha|^2 &= \frac{\cosh (\pi a - \pi b) \cosh (\pi \tilde a + \pi b)}{\cosh (\pi a + \pi b) \cosh (\pi \tilde a - \pi b)}, \label{Nalpha}
\\
|\beta|^2 & = \frac{\sinh (2 \pi b) \sinh (\pi \tilde a - \pi a )}{\cosh (\pi a + \pi b) \cosh (\pi \tilde a - \pi b)}, \label{Nbeta}
\\
\sigma_\mathrm{abs}&  = \frac{\sinh (2 \pi b) \sinh (\pi \tilde a - \pi a )}{\cosh (\pi a - \pi b) \cosh (\pi \tilde a + \pi b)}. \label{Nsigma}
\end{align}

\omits{
The mean number of pairs in near extremal black hole eq. (\ref{Nbeta}) takes the following form
\begin{align} \label{termal interpretation}
 &|\beta|^2  = \frac{e^{2\pi b -2\pi a}-e^{-2\pi b-2\pi a}}{1+e^{-2\pi a-2\pi b}}\  \frac{1-e^{-2\pi \tilde{a}+2\pi a}}{1+e^{-2\pi \tilde{a}+2\pi b}}\cr
&=e^{\frac{m_\mathrm{eff}}{T_\mathrm{RN}}} \! \!  \times \! \underbrace{\frac{e^{-\frac{m_\mathrm{eff}}{T_\mathrm{RN}}} \! \! - \! e^{-\frac{m_\mathrm{eff}}{\mathcal{T}_\mathrm{RN}}}}{1+e^{-\frac{m_\mathrm{eff}}{\mathcal{T}_\mathrm{RN}}}} }_{\substack{\text{Analog Schwinger }\\ \text { effect in  AdS}_2} } \! \! \times \! \underbrace{\frac{e^{-\frac{m_\mathrm{eff}}{T_\mathrm{RN}}}\left(1 \! -e^{-\frac{\o -\tilde{q}A_\tau}{T_H}} \! \right)}{1+e^{-\frac{m_\mathrm{eff}}{T_\mathrm{RN}}}e^{-\frac{\o -\tilde{q}A_\tau}{T_H}}}}_{\substack{\text{Analog Schwinger effect}\\ \text { in  Rindler  space}} },
\end{align}
where $T_H=B/2\pi Q_{\rm s}^2$ is the Hawking temperature. }

\vspace{2em}

\section{Appendix C}
In this Appendix, we present quasinormal modes in the near extremal black hole. It follows from \cite{Chen:2012zn} that the quasinormal mode boundary condition gives the quasinormal mode frequencies
\be
\omega_s = -\frac{b B}{Q_{\rm s}^2}-i \left(\frac 12 + N \right)\frac{B}{Q_{\rm s}^2}.
\ee
Here $N=0, 1, \cdots$. For $B>0$, the quasinormal mode boundary conditions select the negative energy branch of the effective phonon spectrum in the acoustic black hole background.
\end{appendix}

\end{document}